\documentclass[10pt]{iopart}
\usepackage{graphicx}
\usepackage{times}

\usepackage{iopams}         

\newcommand{\M}{\mathcal{M}}                  
\newcommand{\Rest}{\mathcal{R}}               
\newcommand{\Id}{\mathbf{id}}                 
\newcommand{\Norm}[1]{\|#1\|}                 
\newcommand{\bu}{\bi{u}}                      
\newcommand{\bv}{\bi{v}}
\newcommand{\be}{\bi{e}}
\newcommand{\bk}{\bi{k}}
\newcommand{\bg}{\bi{g}}
\newcommand{\bh}{\bi{h}}
\newcommand{\bP}{\bi{P}}
\newcommand{\bQ}{\bi{Q}}
\newcommand{\bT}{\bi{T}}
\newcommand{\etime}{\btau}                       
\newcommand{\espace}{\bsigma}                    
\newcommand{\kbefore}{\bk^{(\mathrm{before})}}   
\newcommand{\kafter}{\bk^{(\mathrm{after})}}     
\newcommand{\PRnabla}{\bnabla_{\dot{\bgamma}}^{\bu}} 

\newcommand{\pubinfo}[1]{                        
   \begin{indented}
   \item[]\rm\raggedright #1
   \end{indented}}

\begin{document}

\title{On Doppler tracking in cosmological spacetimes}

\author{Matteo Carrera and Domenico Giulini}

\smallskip

\address{Physikalisches Institut,
Albert-Ludwigs-Universit\"at,
Hermann-Herder-Stra\ss e 3,\\
D-79104 Freiburg i.~Br.,
Germany}

\eads{carrera@physik.uni-freiburg.de\ \ and\ \ giulini@physik.uni-freiburg.de}

\medskip

\pubinfo{Published in Class.~Quantum Grav.~\textbf{23} (2006) 7483--7492}

\pubinfo{Online at http://stacks.iop.org/CQG/23/7483}

\begin{abstract}
We give a rigorous derivation of the general-relativistic formula for 
the two-way Doppler tracking of a spacecraft in 
Friedmann-Lema\^{\i}tre-Robertson-Walker (FLRW) and in McVittie spacetimes. 
The leading order corrections of the so-determined acceleration to the 
Newtonian acceleration are due to special-relativistic effects and 
cosmological expansion. The latter, although linear in the Hubble constant, 
is negligible in typical applications within the Solar System. 
\end{abstract}

\vspace{-0.5cm}



\section{Introduction}
Doppler measurements are very well studied in the case of weak-field 
approximation of the gravitational field of (arbitrary moving) isolated
sources~\cite{Kopeikin.Schaefer:1999}. However, a similarly careful 
analysis of the cosmological effects on Doppler measurements is, in our 
opinion, still lacking. Considerations in this direction are contained 
in \cite{Laemmerzahl.etal:2006} for FLRW universes and in 
\cite{Kagramanova.etal:2006} for the Schwarzschild-de~Sitter case. 
The aim of our paper is to give a rigorous derivation of the two-way 
Doppler formula in relevant cosmological spacetimes (FLRW and McVittie).

The analysis of the radio Doppler tracking data from the Pioneer\,10
and Pioneer\,11 spacecrafts yield an anomalous inward pointing 
acceleration of magnitude $8.5\times 10^{-10}\,\mathrm{m\ s^{-2}}$; see
\cite{Anderson.etal:2002} and \cite{Markwardt:2002}. This
magnitude is comparable to $Hc=7\times 10^{-10}\,\mathrm{m\ s^{-2}}$, where 
we set $H=72\ \mathrm{km\ s^{-1}\ Mpc^{-1}}$ (see \cite{Spergel.etal:2006} for 
recent figures). This somewhat surprising coincidence invited 
speculations as to a possible cosmological origin of the `Pioneer 
Anomaly'. Whereas there seems to be no disagreement over the absence 
of a genuine \emph{dynamical} influence of cosmological expansion on Solar 
System dynamics, opinions on possible \emph{kinematical} effects are less 
unanimous. Some even seem to claim that the Pioneer Anomaly can be fully
accounted for by such effects~\cite{Rosales.Sanchez-Gomez:1998}
(for a critical discussion, see~\cite{Carrera.Giulini:2006}). 
Note that all these speculations rest on the assumption that the 
cosmological expansion extends into the small region occupied by our 
Solar System, i.e. that the expansion is not screened by local 
inhomogeneities of mass abundance, like that given by our Galaxy and 
further our Solar System. Our results here imply that even in the unlikely 
case that such a screening does not take place, there is no effect due 
to cosmological expansion of the required order of magnitude.

\section{Basic kinematical definitions}
In this section we recall some basic kinematical notions that can be 
found, for example, in~\cite{Bini.etal:1995}.
Let $(\M,\bg,\bnabla)$ represent spacetime, where $\M$ is a four-dimensional%
\footnote{Actually, all formal considerations that follow are independent 
of the dimension of spacetime.} 
manifold (space- and time-orientable) with Lorentzian metric $\bg$ and 
Levi-Civita connection $\bnabla$. 
We adopt the `mostly minus' signature convention where the restriction 
of $\bg$ to spacelike directions is negative definite. We use units in 
which $c=1$. For a vector $\bv$ in the tangent space $T_p\M$ of $\M$ at 
$p\in\M$ we define $\Norm{\bv}_p:=\sqrt{|\bg(\bv,\bv)_p|}$. A 
normalized vector is one for which $\Norm{\bv}=1$.

An \emph{observer (worldline)} is represented by a timelike (smooth) curve 
in $\M$ which w.l.o.g.~we think of as being parameterized with respect to 
arc length and future-directed. By an \emph{observer at} $p\in\M$ we 
understand a timelike future-pointing normalized vector in $T_p\M$. 
An \emph{observer field} is a timelike future-pointing normalized vector 
field on some open subset of $\M$. Let $p$ be a point in $\M$ and $\bu$ 
an observer at $p$. We define two projectors on $T_p\M$ by 
$\bQ_\bu:= \bu \otimes \bu^\flat$, and $\bP_\bu:=\Id-\bQ_\bu$, where 
$\bu^\flat:=\bg(\bu,\cdot)$ is the one-form corresponding to $\bu$ via $\bg$. 
We recall that the projection properties are as follows: $\bP$ and $\bQ$ are 
self-adjoint w.r.t.~$\bg$, $\bQ^2=\bQ$, $\bP^2=\bP$, $\bP\bQ=\bQ\bP=0$, 
and $\bP+\bQ=\Id$.
The \emph{local rest space of $\bu$ at $p$} is then defined as 
$\Rest_\bu(p):=\bP_\bu (T_p\M)$. It locally represents the directions of 
the events that are Einstein-synchronized with the observer $\bu$ at $p$.

The above projectors can be naturally extended to general tensor fields in 
the standard way (e.g.~for 1-forms one defines $\bP\bomega:=\bomega \circ\bP$),
which we will denote by the same symbols. A tensor field $\bT$ is called 
\emph{spatial w.r.t.~$\bu$ (at $p$)} iff $\bP_\bu \bT=\bT$ (at $p$). Given 
an observer $\bu$ at $p$ we split $\bg$ into two degenerate, positive 
semi-definite metrics
\begin{eqnarray}
  &\etime_\bu  &:=  \bQ_\bu \bg \;, \label{eq:time-metric}\\
  &\espace_\bu &:= -\bP_\bu \bg \;, \label{eq:space-metric}
\end{eqnarray}
which measure eigentime- and eigenspace-intervals with respect to $\bu$.

Let now $\bu$ and $\bv$ be two observers at $p$. The \emph{relative velocity 
of $\bv$ w.r.t.~$\bu$} is given by
\begin{equation}\label{eq:relative-velocity-vector}
  \bbeta_\bu(\bv)_p:=\frac{\bP_\bu \bv}{\Norm{\bQ_\bu \bv}} \Big|_p \;,
\end{equation}
and its modulus by
\begin{equation}\label{eq:relative-velocity}
  \beta_\bu(\bv)_p := \Norm{\bbeta_\bu(\bv)}_p =
  \frac{\sqrt{\espace_\bu (\bv,\bv)}}
       {\sqrt{\etime_\bu (\bv,\bv)}}\bigg|_p =
  \sqrt{ 1 - \frac{1}{\bg(\bu,\bv)^2_p} } \;.
\end{equation}
Given a unit vector $\be$ in the rest space of $\bu$ (i.e.~$\bP_\bu \be=\be$ 
and $\Norm{\be}=1$), the relative velocity of $\bv$ w.r.t.~$\bu$ at $p$ in 
direction $\be$ is given by
\begin{equation}\label{eq:relative-velocity-direction}
  \beta_\bu^\be(\bv)_p := 
  \espace_\bu\big( \bbeta_\bu(\bv),\be \big)_p =
  -\bg\big( \bbeta_\bu(\bv),\be \big)_p =
  \frac{-\bg(\be,\bv)}{\bg(\bu,\bv)}\bigg|_p \;.
\end{equation}
Note that the modulus of the relative velocity of $\bv$ w.r.t.~$\bu$ 
equals that of $\bu$ w.r.t.~$\bv$. (Recall that the former velocity is 
measured with clocks and rods at rest w.r.t.~$\bu$ and is represented by 
a vector in $\Rest_\bu(p)$, whereas the latter is measured with clocks and 
rods at rest w.r.t.~$\bv$ and is represented by a vector in $\Rest_\bv(p)$.)  
Clearly $0 \le \beta_\bu(\bv) \le 1$ and $-1 \le \beta_\bu^\be(\bv) \le 1$. 
Moreover $\Norm{\bQ_\bu\bv}=\bg(\bu,\bv)=1/\sqrt{1-\beta_\bu(\bv)^2}$ is 
just the `gamma-factor' familiar from Special Relativity.

The last notion we need to introduce in this section is that of relative
acceleration of a worldline w.r.t.~an observer field. For this purpose, 
given a worldline $\gamma$ and an observer field $\bu$ along it, we define 
\begin{equation}\label{eq:PRnabla}
  \PRnabla:=
  \Norm{\bQ_\bu\dot{\bgamma}}^{-1}
  \bP_\bu \circ \bnabla_{\dot{\bgamma}} \circ \bP_\bu \;.
\end{equation}
$\PRnabla$ is defined on tensor fields along $\gamma$ and $\bP_\bu$ stands
for the extension of the spatial projector on the respective tensor space
(i.e.~every free index is to be projected). On functions we have 
$\PRnabla=\Norm{\bQ_\bu \dot{\bgamma}}^{-1}\dot{\bgamma}$.
Definition (\ref{eq:PRnabla}) resembles that of a rescaled (with the inverse 
of the `gamma-factor' $\Norm{\bQ_\bu\dot{\bgamma}}$) Fermi-derivative along 
$\gamma$. Notice, however, that the projection is taken w.r.t.~$\bu$ and not 
w.r.t.~$\dot{\bgamma}$. It is easy to see that $\PRnabla$ is a derivation on 
spatial (w.r.t.~$\bu$) tensor fields along $\gamma$. Moreover, it can be 
checked that $\PRnabla$ is compatible with the spatial metric, i.e.~the 
metricity property $\PRnabla\,\espace_\bu=0$ holds.

We can now define the \emph{relative acceleration of a worldline 
$\gamma$ w.r.t.~an observer field $\bu$ along $\gamma$} as follows:
\begin{equation}\label{eq:relative-acceleration}
  \balpha_\bu(\gamma) := \PRnabla\,\bbeta_{\bu}(\dot{\bgamma}) =
  \frac{1}{\Norm{\bQ_\bu\dot{\bgamma}}}\bP_\bu \bnabla_{\dot{\bgamma}}
  \bigg(
     \frac{1}{\Norm{\bQ_\bu\dot{\bgamma}}}\bP_\bu \dot{\bgamma}
  \bigg) \;,
\end{equation}
which is a spatial (w.r.t.~$\bu$) vector field along $\gamma$. Clearly, 
the relative acceleration of a worldline $\gamma$ w.r.t.~itself vanishes 
identically, i.e. $\balpha_{\dot{\bgamma}}(\gamma)\equiv 0$, as it should. 
For later use we also note that
\begin{equation}\label{eq:beta-square-derivative}
  \PRnabla\,\beta_\bu(\dot{\bgamma})^2 = 
  2\,\espace_\bu\big( \bbeta_\bu(\dot{\bgamma}), \balpha_\bu(\gamma) \big) \;,
\end{equation}
and
\begin{equation}\label{eq:beta-kappa-derivative}
  \PRnabla\,\beta_\bu^{\be}(\dot{\bgamma}) = 
   \espace_\bu\big( \balpha_\bu(\gamma), \be \big)
  +\espace_\bu\big( \bbeta_\bu(\dot{\bgamma}), \PRnabla\,\be \big) \;,
\end{equation}
for some spatial (w.r.t.~$\bu$) unit vector-field $\be$ along $\gamma$. 
Here we just used the metricity property of $\PRnabla$ and 
$\beta_\bu(\dot{\bgamma})^2 =
\espace_\bu(\bbeta_\bu(\dot{\bgamma}),\bbeta_\bu(\dot{\bgamma}))$.

\section{Light rays}
Consider the propagation of a monochromatic electromagnetic wave. In the 
geometric-optics approximation (i.e.~for wave-lengths negligibly small 
w.r.t.~a typical radius of curvature of the spacetime and w.r.t.~a typical 
length over which amplitude, polarization, and frequency vary) it propagates 
along a lightlike geodesic, along which the wave-vector field, $\bk$, obeys 
$\bg(\bk,\bk)=0$ and $\bnabla_\bk \bk = 0$. 

An observer $\bu$ at $p$ who receives a light signal with wave vector 
$\bk$ measures the frequency
\begin{equation}\label{eq:freqency-measure}
  \omega^{(\bu)}_p(\bk) = \bg(\bu,\bk)_p \;.
\end{equation}
Likewise, a second observer $\bv$ at $p$ will measure the frequency
$\omega^{(\bv)}_p(\bk) = \bg(\bv,\bk)_p$. The relation between the two 
measurements is easily computed using~(\ref{eq:relative-velocity}) 
and~(\ref{eq:relative-velocity-direction}):
\begin{equation}\label{eq:kinematical-redshift}
  \frac{\omega^{(\bv)}_p(\bk)}{\omega^{(\bu)}_p(\bk)} =
  \frac{1}{\sqrt{1-\beta_\bu(\bv)^2}} \Big( 1-\beta_\bu^{\hat \bk}(\bv) \Big)
  \Big|_p\;,
\end{equation}
where $\hat \bk:=\bP_\bu\bk/\Norm{\bP_\bu\bk}$ is the normalized vector in 
the rest space of $\bu$ at $p$ that points in the direction of the light 
propagation. This is nothing but the general Doppler formula. 

Let the observer $\bv$ at $p$ carry a mirror (meaning that the mirror is at 
rest w.r.t.~$\bv$) that is used to \emph{reflect back} the light ray. This 
means that the frequency, as measured by $\bv$, does not change whereas the 
spatial projection of the wave vector $\bk$ changes sign. In short, the 
process of reflection is given by 
\begin{equation}\label{eq:reflection}
  \kafter|_p = \bQ_\bv \kbefore|_p - \bP_\bv \kbefore|_p \;.
\end{equation}
Another observer, say $\bu$, at $p$ will see a frequency shift according 
to 
\begin{equation}\label{eq:frequency-reflection-general}
  \frac{\omega^{(\bu)}_p(\kafter)}{\omega^{(\bu)}_p(\kbefore)} =
  2 \, \frac{1-\beta_\bu^{\hat \bk}(\bv)_p}{1-\beta_\bu(\bv)^2_p} - 1 \;,
\end{equation}
as one can compute from~(\ref{eq:reflection}), (\ref{eq:kinematical-redshift}),
and~(\ref{eq:relative-velocity}). Note that here and henceforth the wave 
vector $\bk$ that defines $\hat \bk$ at the reflection point $p$ refers to 
$\bk$ just \emph{before} reflection.

\section{Two-way Doppler tracking in a homogeneous and isotropic universe}
We consider a FLRW cosmological model, given by the metric 
\begin{equation}\label{eq:FLRW-metric}
  \bg = \rmd t^2 - a^2(t)\,\bh \;,
\end{equation}
together with the geodesic cosmological observer-field
\begin{equation}\label{eq:cosmological-vectorfield}
  \bu=\partial/\partial t \;.
\end{equation}
In~(\ref{eq:FLRW-metric}) $\bh$ is a homogeneous and isotropic metric (hence 
of constant curvature) on the slices of constant cosmological time~$t$. 

\begin{figure}[t]
\centering
\includegraphics[width=4.5truecm]{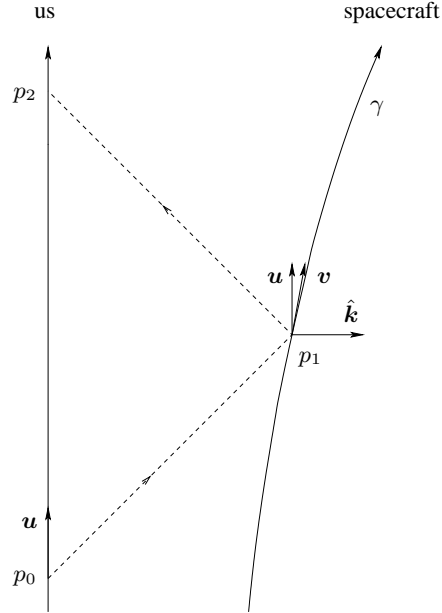}
\put(-132,225){\small us}
\put( -15,225){\small spacecraft}
\put(-140, 12){\small $p_0$}
\put(-137, 32){\small $\bu$}
\put( -32, 95){\small $p_1$}
\put( -44,125){\small $\bu$}
\put( -25,125){\small $\bv$}
\put( -15,110){\small $\hat \bk$}
\put(-140,195){\small $p_2$}
\put(  -5,190){\small $\gamma$}
\caption{\label{fig:Doppler-tracking}
Sketch of the two-way Doppler tracking: The observer (us) moves on an 
integral curve of the cosmological vector field $\bu$. At $p_0$ he sends an 
electromagnetic signal to the spacecraft which moves along the worldline 
$\gamma$. The signal is then reflected back from the spacecraft at $p_1$ 
and, finally, received at $p_2$ by the observer. 
$\hat \bk:=\bP_\bu \bk/\Norm{\bP_\bu \bk}$ is the normalized spatial 
(w.r.t.~$\bu$) wave vector of the infalling light ray.}
\end{figure}

Let $\bk$ be the wave-vector field along a light ray (worldline). The 
frequency, as measured by the cosmological observers $\bu$, will change 
along the ray according to the well-known cosmological red-shift relation 
(see e.g.~\cite{Straumann:2004}) 
\begin{equation}\label{eq:cosmological-redshift}
  \omega^{(\bu)}(\bk) \; (a\circ t) = \mbox{constant along the light ray} \;.
\end{equation}
The two-way Doppler tracking (see figure~\ref{fig:Doppler-tracking}) now 
consists in the exchange of light (radar) signals between us, the observers 
on Earth who hypothetically moves along the cosmological flow of $\bu$, and 
another observer $\gamma$ (spacecraft). 
Schematically the tracking involves the following five processes: 
  (i)~emission of the signal at $p_0$,
 (ii)~propagation from $p_0$ to $p_1$,
(iii)~reflection at $p_1$,
 (iv)~propagation from $p_1$ to $p_2$, and, finally,
  (v)~reception at $p_2$.
Accordingly, using the cosmological red-shift relation 
(\ref{eq:cosmological-redshift}) between $p_0$ and $p_1$, the reflection 
shift (\ref{eq:frequency-reflection-general}) at $p_1$, and again the 
cosmological red-shift relation (\ref{eq:cosmological-redshift}) between 
$p_1$ and $p_2$, one gets
\begin{equation}\label{eq:Doppler-tracking}
  \frac{ \omega^{(\bu)}_{p_2}(\bk) }{ \omega^{(\bu)}_{p_0}(\bk) } = 
  \frac{ a(t(p_0)) }{ a(t(p_2)) }
  \left(
    2\,\frac{1-\beta_\bu^{\hat \bk}(\bv)_{p_1}}{1-\beta_\bu(\bv)^2_{p_1}}-1
  \right)\;.
\end{equation}

This is the formula for the two-way Doppler tracking in a FLRW spacetime.
It relates the spacecraft's velocity (relative to $\bu$) with the observable 
quantities (emitted and received frequencies, emission and reception times) 
if the scale function $a$ is known. To get a better feeling of the above 
relation we note that to linear order in the two quantities $H \Delta t_{20}$ 
and $\beta$ it reads
\begin{equation}\label{eq:Doppler-tracking-approx}
  \frac{ \omega^{(\bu)}_{p_2}(\bk) }{ \omega^{(\bu)}_{p_0}(\bk) }
  \approx  1 - 2\beta_{\bu}^{\hat\bk}(\bv)_{p_1} - H\Delta t_{20} \;,
\end{equation}
where $\Delta t_{20}:=t(p_2) - t(p_0)$ is the coordinate-time interval 
between the emission and the reception events, which, since we are moving
along $\partial/\partial t$, equals the eigentime interval measured by us 
between the two events. $H:=\dot a/a$ denotes the Hubble parameter, which 
in (\ref{eq:Doppler-tracking-approx}) can be evaluated at any of the times 
in the interval $[t_0, t_2]$.

Now, while tracking a spacecraft by continuously emitting a radar signal 
with constant frequency $\omega_0$ the whole construction is pushed forward 
in time. One should think of the worldlines of the spacecraft ($\gamma$) 
and ours as given. The latter is taken to be an integral curve of the 
cosmological observer-field $\bu=\partial/\partial t$ and the former is 
given by some equation of motion (plus initial conditions), which we do 
not need to specify here. Thus, the three points (events) $p_0$, $p_1$, and 
$p_2$ are uniquely determined by any one of them. The same holds for the 
respective times $t_i:=t(p_i)$. We can then choose to express the events 
$p_i$ as functions of the reception time $t_2$. Doing this, the reflection 
time $t_1$ and the emission time $t_0$ become functions of $t_2$ as well.

One of the major tasks for the two-way Doppler tracking is to determine 
the spacecraft's (spatial) acceleration. Since (\ref{eq:Doppler-tracking}) 
relates the frequency shift with the velocity, differentiation of 
(\ref{eq:Doppler-tracking}) w.r.t.~the reception time gives a relation
between the frequency-shift \emph{rate} on one hand, and the acceleration 
and velocity on the other. 
In the differentiation w.r.t.~$t_2$, one has to take care of the different
time dependences of the quantities in question. For example, the received 
frequency is to be thought of as function of the reception time $t_2$ and 
the spacecraft's velocity as a function of the reflection time $t_1$,
which, in turn, is to be thought of as function of $t_2$. 
In order to differentiate (\ref{eq:Doppler-tracking}) w.r.t.~$t_2$ we thus 
need to know the dependences of the emission and reflection times, $t_0$ 
and $t_1$, on the reception time $t_2$. These dependences can be obtained 
from the fact that the pairs of events $(p_0,p_1)$ and $(p_1,p_2)$ are 
lightlike separated. For simplicity, we will specialize to the spatially 
flat case, which is compatible with current observational 
data~\cite{Spergel.etal:2006}. Introducing spherical coordinates,
$(r,\theta,\varphi)$, on the slices of constant $t$, the spatial metric reads 
$\bh = \rmd r^2 + r^2(\rmd\theta^2+\sin^2\theta\ \rmd\varphi^2)$.
Without loss of generality we may assume our worldline to be given by 
$r=0$. The spacecraft's worldline is described by some functions 
$(t_1,r_1,\theta_1,\varphi_1)$ of the reflection 
time $t_1$. Hence one obtains
\begin{equation}\label{eq:reflection-time}
  \int_{t_1(t_2)}^{t_2} \frac{\rmd t}{a(t)} = 
  -\frac{1}{c}\int_{r_1(t_1(t_2))}^{r_2} \rmd r \;,
\end{equation}
where the minus sign is because the light ray is `inward' pointing.
Differentiating (\ref{eq:reflection-time}) w.r.t.~$t_2$ and noting that
$\beta_\bu^{\hat\bk}(\bv)_{p_1} = a(t_1)\dot{r}_1/\dot{t_1} = 
a(t_1)\rmd r_1/\rmd t_1$ (where 
$\hat{\bk}=\Norm{\partial/\partial r}^{-1}\partial/\partial r$ and the 
overdot denotes differentiation w.r.t.~the eigentime of $\gamma$), one gets
\begin{equation}\label{eq:reflection-time-derivative}
  \frac{\rmd t_1}{\rmd t_2} = \frac{a(t_1)}{a(t_2)}
  \Big( 1+\beta_\bu^{\hat\bk}(\bv)_{p_1} \Big)^{-1} \;.
\end{equation}
Similarly, differentiating
\begin{equation}\label{eq:emission-time}
  \int_{t_0(t_2)}^{t_2} \frac{\rmd t}{a(t)} = 
  +\frac{1}{c}\int_{r_0}^{r_1(t_1(t_2))} \rmd r 
  -\frac{1}{c}\int_{r_1(t_1(t_2))}^{r_2} \rmd r 
\end{equation}
w.r.t.~$t_2$ and using (\ref{eq:reflection-time-derivative}) one gets
\begin{equation}\label{eq:emission-time-derivative}
  \frac{\rmd t_0}{\rmd t_2} = \frac{a(t_0)}{a(t_2)}
  \bigg( \frac{ 1 - \beta_\bu^{\hat\bk}(\bv)_{p_1} }
              { 1 + \beta_\bu^{\hat\bk}(\bv)_{p_1} } \bigg) \;.
\end{equation}

In passing we remark that (\ref{eq:emission-time-derivative}) can be used 
to compute the variation of the light round-trip-time $\Delta t_{20}$ as 
measured by the receiver ($\Delta t_{20}$ is just twice the radar distance 
between observer and spacecraft):
\begin{equation}\label{eq:light-time-variation}
  \frac{\rmd}{\rmd t_2} \Delta t_{20} =
  1 - \frac{a(t_0)}{a(t_2)}
  \bigg( \frac{ 1 - \beta_\bu^{\hat\bk}(\bv)_{p_1} }
              { 1 + \beta_\bu^{\hat\bk}(\bv)_{p_1} } \bigg)
  \approx 2\,\beta_\bu^{\hat\bk}(\bv)_{p_1} + H\Delta t_{20} \;,
\end{equation}
which clearly displays the expected contributions due to the spacecraft's 
motion and the cosmological expansion respectively.

Returning to the derivation of the frequency-shift rate, we notice that 
$\rmd/\rmd t_1 = (\rmd t_1/\rmd\tau_1)^{-1}\rmd/\rmd\tau_1=\PRnabla$
on functions along the spacecraft's worldline $\gamma$, where $\tau_1$ 
is its arc-length. Taking into account (\ref{eq:beta-square-derivative}), 
(\ref{eq:beta-kappa-derivative}), (\ref{eq:reflection-time-derivative}), 
and (\ref{eq:emission-time-derivative}), differentiation of 
(\ref{eq:Doppler-tracking}) w.r.t.~$t_2$ gives
\begin{eqnarray}\label{eq:Doppler-shift-rate}
\fl
  \frac{1}{\omega_0}\frac{\rmd\omega_2}{\rmd t_2} =
   -\,\frac{a(t_0)}{a(t_2)}
  \Bigg\{\,
    &2\Big[ \espace(\balpha,\hat{\bk})+\espace(\bbeta,\PRnabla\hat{\bk}) \Big]
    \frac{a(t_1)}{a(t_2)}
    \Big( 1+\beta^{\hat{\bk}} \Big)^{-1} \Big( 1-\beta^2 \Big)^{-1}
    \nonumber\\
    &-4\,\espace(\balpha,\bbeta)\,
    \frac{a(t_1)}{a(t_2)}
    \bigg( \frac{ 1-\beta^{\hat{\bk}} }{ 1+\beta^{\hat{\bk}} } \bigg)
    \Big( 1-\beta^2 \Big)^{-2}
    \nonumber\\
    &+
    \Bigg( 
        \frac{\dot a(t_2)}{a(t_2)}
        -\frac{\dot a(t_0)}{a(t_2)}
        \bigg(\frac{1-\beta^{\hat{\bk}}}{1+\beta^{\hat{\bk}}} \bigg)
    \Bigg)
    \bigg(
        \frac{1-2\beta^{\hat{\bk}}+\beta^2}{1-\beta^2}
    \bigg)
  \Bigg\} \;.
\end{eqnarray}
Here, we suppressed the arguments and the indices $\bu$ for the sake of
readability and put $\omega_2:=\omega_{p_2}^{(\bu)}(\bk)$. This formula 
gives the exact relation between the observable frequency-shift rate 
(measured `here') and the local kinematical properties of the spacecraft 
(defined `there') in a spatially flat FLRW spacetime---provided the scale 
function $a$ is known.

In the special, but relevant, case where the spacecraft's motion is in 
direction of the line-of-sight (`radial' for short), meaning that both 
$\bbeta$ and $\balpha$ are collinear to $\hat{\bk}$, we have 
$\bbeta=\beta^{\hat{\bk}}\,\hat{\bk}$, and
$\balpha=\alpha^{\hat{\bk}}\,\hat{\bk}$,
where $\alpha^{\hat{\bk}}:=\espace(\balpha,\hat{\bk})$. Moreover, it is 
straightforward to check that $\PRnabla\hat{\bk}=0$.
The two-way Doppler-tracking formula (\ref{eq:Doppler-shift-rate}) 
now simplifies to
\begin{eqnarray}\label{eq:Doppler-tracking-radial}
\fl
  \frac{1}{\omega_0}\frac{\rmd\omega_2}{\rmd t_2} =
  -\,\frac{a(t_0)}{a(t_2)}\frac{a(t_1)}{a(t_2)}
  \Bigg\{\,
    &2\,\alpha^{\hat{\bk}}\,( 1+\beta^{\hat{\bk}} )^{-3}
    \nonumber\\
    &+\Bigg( \frac{\dot a(t_2)}{a(t_1)}
          -\frac{\dot a(t_0)}{a(t_1)}
           \bigg( \frac{1-\beta^{\hat{\bk}}}{1+\beta^{\hat{\bk}}} \bigg)
    \Bigg) \bigg( \frac{1-\beta^{\hat{\bk}}}{1+\beta^{\hat{\bk}}} \bigg)
  \Bigg\} \;.
\end{eqnarray}
This formula is still exact if one restricts to radially moving spacecrafts. 
In a linear approximation in $H\Delta t_{20}$ it becomes 
\begin{equation}\label{eq:Doppler-tracking-radial-approx}
  \frac{1}{\omega_0}\frac{\rmd\omega_2}{\rmd t_2} \approx
  -2 \bigg( 1\!-\!3 H\frac{\Delta t_{20}}{2} \bigg) 
  \Bigg\{ 
     \alpha^{\hat{\bk}} ( 1+\beta^{\hat{\bk}} )^{-3} 
     +H\beta^{\hat{\bk}}  
     \frac{ 1-\beta^{\hat{\bk}} }{ (1+\beta^{\hat{\bk}})^2 }
  \Bigg\} \,.
\end{equation}
In all cases of interest one is also interested in a slow-motion 
approximation involving an expansion in $\beta$. Retaining only terms 
linear in $\beta$ and $H\Delta t$ we simply get 
$(\rmd \omega_2/\rmd t_2)/\omega_0 \approx -2\,\alpha^{\hat{\bk}}$.
This is exactly the result one obtains from Newtonian physics. Leading 
order corrections are obtained by going to quadratic order in $\beta$ 
and also keeping mixed terms $\beta\!\cdot\!H\Delta t$. The result is
\begin{equation}\label{eq:Doppler-tracking-radial-quadratic}
  \frac{1}{\omega_0}\frac{\rmd\omega_2}{\rmd t_2} \approx
  -2\,
  \Big\{ 
      \alpha^{\hat{\bk}}\,\Big( 1-3\beta^{\hat{\bk}}-3H\Delta t_{20}/2 \Big)
      +H\beta^{\hat{\bk}}
  \Big\} \;.
\end{equation}

At this point it is crucial to understand how $\balpha$ relates to the 
acceleration that appears on the left-hand side of Newton's equation of 
motion. The latter is derived as the weak-field and slow-motion 
approximation of the spatial part of the geodesic equation. Recall that 
`spatial part' refers to the choice of the observer field, from which 
the establishment of a Newtonian equation always depends. 

In addition, in order to write down a Newtonian equation, we need to use 
spatial coordinates of direct metrical significance. This is achieved by 
using a rescaled radial coordinate, $r_*(r,t):=a(t)r$ such that $r_*$ is 
now the proper geodesic distance to the origin (us) within each time slice 
of constant $t$. The time coordinate remains unchanged, $t_*:=t$, so that 
\begin{equation}
\label{eq:DbyDtStar}
  \partial/\partial t_* = \partial/\partial t - H(t)r\;\partial/\partial r\;.
\end{equation}
The observer field is now the normalization of that field:
\begin{equation}\label{eq:Newtonian-observer-field}
  \bu_* = \Norm{\partial/\partial t_*}^{-1}\partial/\partial t_*\;.
\end{equation}
It also results from applying a (spacetime dependent) boost of 
parameter $-\beta_\bu(\bu_*)=-Hr_*$ in the local planes spanned by 
$\bu$ and $\be_r:=\Norm{\partial/\partial r}^{-1}\partial/\partial r$.
Note also that $\bu_*$ is hypersurface orthogonal, though clearly not 
orthogonal to the slices of constant $t$.

The spatial projection of the geodesic equation with respect to $\bu_*$
up to quadratic order in $H \Delta t$ and linear order in 
$\beta_{\bu_*}(\dot{\bgamma})$ turns out to be
\begin{equation}\label{eq:cosmo-Newton-eq}
  \balpha_{\bu_*}(\gamma) \approx 
  \bigg( \, \frac{\ddot a}{a}\, r_* \bigg) \;\be_r \circ \gamma\;.
\end{equation}
This is the form of the equation of motion in which dynamical 
considerations are most conveniently 
addressed~(see \cite{Carrera.Giulini:2006} and references therein). 
Hence in the above formulae for the Doppler tracking we need to express 
the relative accelerations and velocities defined w.r.t.~$\bu$ in terms 
of relative accelerations and velocities w.r.t.~$\bu_*$. In the FLRW case 
these transformations are given in \cite{Carrera.Giulini:2006}. For a 
general spacetime they can be found e.g.~in~\cite{Bini.etal:1995}. 
However, for our discussion it is sufficient to consider the approximated 
Doppler tracking formula (\ref{eq:Doppler-tracking-radial-quadratic}) and 
within its level of approximation it makes no difference whether we refer 
spatial acceleration and velocities w.r.t.~$\bu$ or w.r.t.~$\bu_*$.
Hence (\ref{eq:Doppler-tracking-radial-quadratic}) gives the sought-after
relation between the two-way Doppler measurement and the kinematical 
quantities that enter the Newtonian equation.

Looking at (\ref{eq:Doppler-tracking-radial-quadratic}) we see, among other 
corrections, an additional acceleration term $Hc\beta^{\hat{\bk}}$ 
(reintroducing the speed of light $c$). Hence the sometimes alleged $Hc$ 
acceleration term~\cite{Rosales.Sanchez-Gomez:1998} is actually 
suppressed by a factor $\beta$, as already pointed out 
in~\cite{Laemmerzahl.etal:2006}, though we cannot agree with the derivation 
given there. In particular we conclude that, even if the universe did expand 
down to scales of the Solar System, there would be no effect of the same 
order of magnitude as the Pioneer Anomaly. Besides, notice that the above 
mentioned correction points in outer direction and hence opposite to the 
Pioneer anomalous acceleration. One expects that this conclusion remains 
true in the presence of an isolated local inhomogeneity, such as a single 
star, since the contributions coming from the `gravitational' red-shift 
of the star cancel out in the two-way process. Insofar as such an 
inhomogeneity can be modeled by the McVittie metric this is explicitly 
shown in the next section.

\section{Two-way Doppler tracking in a McVittie spacetime}
The so-called `flat' McVittie model with central mass $m_0$ and 
scale-factor $a(t)$ is given by the spacetime metric
\begin{equation}\label{eq:McVittie-metric}
  \bg=\left( \frac{1-m_0/2a(t)r}{1+m_0/2a(t)r} \right)^2\!\rmd t^2
     -\left( 1 + \frac{m_0}{2a(t)r} \right)^4\!a^2(t)
     (\rmd r^2\!+\!r^2\rmd \Omega^2)
\end{equation}
together with a cosmological observer-field
\begin{equation}\label{eq:cosmological-observer-McV}
  \bu=\Norm{\partial/\partial t}^{-1}\partial/\partial t
\end{equation}
along which the cosmological matter, given by an ideal fluid with 
pressure, moves.

The red-shift formula for a radial light ray in a McVittie spacetime is
easily computed up to linear order in $H\Delta t_{10}$ and $m_0/ar$ 
to be
\begin{equation}\label{eq:McVittie-redshift}
  \frac{\big(\,\Norm{\partial/\partial t}\,\omega^{(\bu)} \big)_{p_1}}
       {\big(\,\Norm{\partial/\partial t}\,\omega^{(\bu)} \big)_{p_0}}
  \approx 1 - H\Delta t_{10} \;,
\end{equation}
where $\Delta t_{10}:=t(p_1)-t(p_0)$ and $H:=\dot a/a$ as previously.
In case $H=0$ this reduces to the familiar gravitational red-shift 
relation in Schwarzschild spacetime, as it should. 
By (\ref{eq:McVittie-redshift}) one sees that the main formula 
(\ref{eq:Doppler-tracking}) for the two-way Doppler 
frequency-shift we derived in the FLRW case still remains valid for the 
McVittie case in linear order in $H\Delta t$ and $m_0/ar$. The reason 
is simply that the gravitational contribution to the red-shift (coming 
from the factors $\Norm{\partial/\partial t}\approx 1-m_0/ar$ in 
(\ref{eq:McVittie-redshift})) vanishes in the two-way process within the 
considered approximation.

As in the previous section, we wish to compute the Doppler frequency-shift 
rate measured by an observer (us) along $\bu$ who exchanges light (radar) 
signals with a spacecraft moving along an arbitrary worldline $\gamma$ 
(see figure \ref{fig:Doppler-tracking}). For this we need the generally 
valid relations (\ref{eq:beta-square-derivative}) and 
(\ref{eq:beta-kappa-derivative}) and the relations 
(\ref{eq:reflection-time-derivative}) and 
(\ref{eq:emission-time-derivative}) which, fortunately, hold exactly also 
in the McVittie case, provided that central mass, observer, and spacecraft 
are aligned. The latter will be assumed henceforth.

A difference to the FLRW case is that here $t$ is no longer the 
eigentime along the observer's worldline. Rather, to get the measured 
frequency-shift rate, we have to differentiate (\ref{eq:Doppler-tracking}) 
with $\bu=\rmd/\rmd \tau_2=\Norm{\partial/\partial t}^{-1}\rmd/\rmd t_2$. 
When differentiating the kinematical factor on the r.h.s.~of 
(\ref{eq:Doppler-tracking}) we proceed as before but now take 
into account that $\rmd/\rmd t_1=(\rmd t_1/\rmd\tau_1)^{-1}\rmd/\rmd\tau_1=
\Norm{\partial/\partial t}\,\PRnabla$ on functions along the spacecraft's 
worldline $\gamma$, where $\tau_1$ is its arc-length. This leads to the 
formula
\begin{equation}\label{eq:McVittie-Doppler-tracking}
  \frac{1}{\omega_0}\frac{\rmd\omega_2}{\rmd \tau_2} \approx
  -2\,
  \bigg\{\, 
      \alpha^{\hat{\bk}}\,
      \bigg( 1 - 3\beta^{\hat{\bk}} - 3H\frac{\Delta \tau_{20}}{2} 
              + \frac{m_0}{R^2}\frac{\Delta\tau_{20}}{2} \bigg)
      +H\beta^{\hat{\bk}}\,
  \bigg\}
\end{equation}
which is valid in linear order in $H\Delta t_{20}$ and $m_0/ar$ and 
quadratic order in $\beta$, and differs from 
(\ref{eq:Doppler-tracking-radial-quadratic}) merely in the term 
containing $m_0$ (note that in (\ref{eq:Doppler-tracking-radial-quadratic})
$\Delta t_{20}=\Delta\tau_{20}$). Formula 
(\ref{eq:McVittie-Doppler-tracking}) shows that in the McVittie case, 
too, the acceleration correction term $Hc$ is suppressed by a 
factor $\beta$.

As for the FLRW case, we may express the relative acceleration and velocity 
defined w.r.t.~$\bu$ in terms of the relative acceleration and velocity 
w.r.t.~$\bu_*$. As previously, $\bu_*$ is the normalized future-directed 
vector field whose integral lines have constant proper radius $r_*$ 
(i.e.~radial geodesic distance%
\footnote{In~\cite{Carrera.Giulini:2006}, for the McVittie case, we take 
the areal radius as radial coordinate for the Newtonian equation just for 
computational convenience. In the considered approximation, and in the 
region of validity of the Newtonian equation, the areal radius equals the 
geodesic radial distance.}
on the slices of constant $t$) and constant angular coordinates.

Up to now $\gamma$ was unspecified. Let us now assume $\gamma$ to be 
a geodesic. In leading order in $H\Delta\tau$ and $m_0/r_*$ and 
linear order in $\beta_{\bu_*}(\dot{\bgamma})$ the geodesic equation 
then leads to the following expression for the spatial acceleration 
w.r.t.~$\bu_*$:
\begin{equation}\label{eq:McV-Newton-eq}
  \balpha_{\bu_*}(\gamma) \approx 
  \bigg( \frac{\ddot a}{a}\, r_* - \frac{m_0}{r_*^2} \bigg) \;\be_r 
  \circ \gamma\;.
\end{equation}
Here we made use of the identity $\ddot a/a=-qH^2$, where $q$ is the 
deceleration parameter, which shows that $\ddot a/a$ is of quadratic 
order in $H$, given that $q$ is not a large number. 
Formula (\ref{eq:McV-Newton-eq}) is the form of the equation of 
motion in the next-to-Newtonian approximation in which dynamical 
considerations are most conveniently addressed~(see again 
\cite{Carrera.Giulini:2006} and references therein).

In the special case of purely radial motion, insertion of 
(\ref{eq:McV-Newton-eq}) into (\ref{eq:McVittie-Doppler-tracking})
gives a formula predicting the two-way Doppler-shift rate in linear 
order in $H\Delta\tau_{20}$ and $m_0/r_*$ and quadratic order in 
$\beta_{\bu_*}(\dot{\bgamma})$:
\begin{equation}\label{eq:McVittie-Doppler-tracking-geodesic}
  \frac{1}{\omega_0}\frac{\rmd\omega_2}{\rmd \tau_2} \approx
  -2\,
  \bigg\{ -\frac{m_0}{r_*^2} \Big( 1 - 3\beta^{\hat{\bk}} \Big)
          +H\beta^{\hat{\bk}}\,
  \bigg\} \;.
\end{equation}
Hence there are two corrections to the Newtonian contribution. One is 
proportional to $H$ and stems from the cosmological expansion, the other 
is independent of $H$ and of purely special-relativistic origin, as one 
might check explicitly. Their ratio is (up to a factor $\sqrt{3}$) given 
by the square of the ratio of $r_*$ to the geometric mean of the 
Schwarzschild radius $m_0$ and the Hubble radius $1/H$. 
The latter is of the order of $10^{23}\,\mathrm{km}$, so that its geometric 
mean with a Schwarzschild radius of $1\ \mathrm{km}$ is approximately given 
by $2400\ \mathrm{AU}$. The ratio of the effects is therefore of the 
order $10^{-7}$. Hence the cosmological contribution is negligible for any 
application in the Solar System as compared to the special-relativistic 
correction. For the Pioneer spacecrafts 10 and 11 we have a radial velocity 
of about $12\ \mathrm{km\ s^{-1}}$. This amounts to a special relativistic 
correction of magnitude $10^{-4}$ times the Newtonian gravitational 
acceleration, in an outwardly pointing direction. This is of the same order 
of magnitude as the Pioneers' anomalous 
acceleration~\cite{Anderson.etal:2002}, but pointing in the opposite 
direction.

\section{Conclusion}
It is often heard that effects of cosmological expansion are at most of 
order $H^2$. We have seen in (\ref{eq:McV-Newton-eq}) that this is true 
as far as deviations from the Newtonian equation of motion are concerned. 
However, there is a term linear in $H$ that enters the formula for the 
two-way Doppler tracking, as seen in 
(\ref{eq:McVittie-Doppler-tracking-geodesic}), even though the correction 
it gives is negligible as compared with the special relativistic one, due 
to its suppression with a factor of $\beta$. 

Finally we point out that for direct comparison with actual measurements 
there are other corrections to our Doppler formulae that can easily be 
taken care of. For example, one needs to incorporate the motion of the 
Earth relative to the cosmological substrate with an additional Doppler 
factor (\ref{eq:kinematical-redshift}) at the emission and reception points 
$p_0$ and $p_2$. Also, one needs to take into account that in the actual 
tracking procedure signals are sent back with a fixed frequency translation 
factor (see e.g.~formula~(1) in~\cite{Markwardt:2002}).

\end{document}